%% file: main.tex
\pdfoutput=1
\documentclass{article}
\usepackage[preprint]{spconfa4,amsmath,graphicx}

\usepackage{multirow}
\usepackage{pifont}
\usepackage{booktabs}
\usepackage{amssymb}
\usepackage{amsmath}
\usepackage{tikz}
\usepackage{enumitem}
\usepackage{xspace}
\usepackage{pgfplots}\usepackage[acronym, toc,nonumberlist]{glossaries}
\usepackage{balance}
\usepackage{csquotes}
\usepackage{hyperref}
\usepackage{url}
\usepackage[capitalize]{cleveref}

\usepackage{siunitx}
\robustify\bfseries  
\usepackage{amsmath,graphicx}
\usetikzlibrary{calc}
\usetikzlibrary{tikzmark}
\usetikzlibrary{arrows, positioning}
\usetikzlibrary{fit}
\usepgfplotslibrary[groupplots]
\pgfplotsset{compat=newest}

\newcommand{\cmark}{\text{\ding{51}}}
\newcommand{\xmark}{\text{\ding{55}}}

\makeatletter
\def\wrt{w.r.t.\@\xspace}
\makeatother

\newglossaryentry{WSJ0-2mix}{name={WSJ0-2mix}, description={}}
\newglossaryentry{WSJ0-3mix}{name={WSJ0-3mix}, description={}}
\newglossaryentry{PIT-BLSTM}{name={PIT-BLSTM}, description={}}

\newacronym{ASR}{ASR}{Automatic Speech Recognition}
\newacronym{CTFA}{CTFA}{Convolutive Transfer Function Approximation}
\newacronym{MTFA}{MTFA}{Multiplicative Transfer Function Approximation}
\newacronym{PIT}{PIT}{Permutation Invariant Training}
\newacronym{RIR}{RIR}{Room Impulse Response}
\newacronym{STFT}{STFT}{Short-Time Fourier Transform}
\newacronym{siSDR}{SI-SDR}{scale-invariant Signal-to-Distortion Ratio}
\newacronym{SDR}{SDR}{Signal-to-Distortion Ratio}
\newacronym{SRMR}{SRMR}{Signal-to-Reverberation Modulation energy Ratio}
\newacronym{WER}{WER}{Word Error Rate}
\newacronym{WDO}{WDO}{W-Disjoint Orthogonality}
\newacronym{WPE}{WPE}{Weighted Prediction Error}
\newacronym{DM}{DM}{Dynamic Mixing}

\copyrightnotice{978-1-6654-6867-1/22/\$31.00~\copyright2022 IEEE}
                   
\title{Monaural source separation: \\From anechoic to reverberant environments}
\name{\begin{tabular}{c}Tobias Cord-Landwehr$^{\star}$, Christoph Boeddeker$^{\star}$, Thilo von Neumann$^{\star}$, \\ C\u{a}t\u{a}lin Zoril\u{a}$^{\dagger}$, Rama Doddipatla$^{\dagger}$, Reinhold Haeb-Umbach$^{\star}$\end{tabular}}
\address{$^{\star}$ Paderborn University, Department of Communications Engineering, Paderborn, Germany \\
$^{\dagger}$ Toshiba Cambridge Research Laboratory, Cambridge, United Kingdom}
\begin{document}
\ninept
\maketitle
\begin{abstract}
Impressive progress in neural network-based single-channel speech source separation has been made in recent years.
But those improvements have been mostly reported on anechoic data, a situation that is hardly met in practice.
Taking the SepFormer as a starting point, which achieves state-of-the-art performance on anechoic mixtures, we gradually modify it to optimize its performance on reverberant mixtures.
Although this leads to a word error rate improvement by 7 percentage points compared to the standard SepFormer implementation, the system ends up with only marginally better performance than a \gls{PIT-BLSTM} separation system, that is optimized with rather straightforward means. This is surprising and at the same time sobering, challenging the practical usefulness of many improvements reported in recent years for monaural source separation on nonreverberant data.
\end{abstract}
\section{Introduction}
\label{sec:intro}




Neural network-based single-channel source separation has made significant advances in the last years.
Starting with the seminal papers on deep clustering \cite{16_Hershey_dc_wsjmix} and \gls{PIT} \cite{Kolbaek_17_uPit}, improvements have been achieved by combining the two in a multi-objective training criterion \cite{17_Luo_chimera}, or replacing the \gls{STFT} with a learnable encoder and decoder 
\cite{18_Luo_tasnet}.
Employing convolutional mask estimation network architectures \cite{19_Luo_convtasnet} or accounting for short- and longer-term correlations in the signal with recurrent network layers \cite{20_Luo_dprnn} and combining them with a transformer architecture \cite{Subakan_21_sepformer} further elevated the performance.
Overall, this has led to an improvement in \gls{siSDR} from roughly \SI{10}{dB} to more than \SI{20}{dB} on the standard \gls{WSJ0-2mix} data set \cite{16_Hershey_dc_wsjmix}, which consists of artificial mixtures of nonreverberant speech.\footnote{\url{https://paperswithcode.com/sota/speech-separation-on-wsj0-2mix}}

However, an anechoic environment is a rather unrealistic assumption for speech separation as
in a real-world scenario, the superposition of the speech of two or more speakers typically occurs in a distant microphone setting.
A distant microphone naturally captures a reverberated signal. A practically much more relevant setting is thus the separation of mixtures of reverberated speech.

Source separation of noisy and  reverberant mixtures is much harder.
In particular, reverberation has been considered more challenging than noise \cite{20_Maciejewski_WHAMR}.
This comes to no surprise because the key assumptions underlying monaural mask-based source separation, namely the sparsity and orthogonality of speech representations in the \gls{STFT} domain, tend to break down under reverberation.

WHAMR! \cite{20_Maciejewski_WHAMR} and SMS-WSJ \cite{Drude_19_smswsj} are two widely used data sets for research on source separation for reverberant mixtures.
Both contain artificially reverberated utterances from the WSJ corpus.
While WHAMR! 
additionally contains environmental noise, SMS-WSJ consists of 6-channel microphone array data and allows for performance comparison \wrt \gls{WER} as it is accompanied by a Kaldi recipe \cite{11_Povey_kaldi}. 
Source separation performance on WHAMR! is in the range of \SIrange{2}{8}{\decibel} output \gls{siSDR} \footnote{Obtained by comparing the reported improvement with the input \gls{siSDR} of \SI{-6}{\decibel}}, while the performance on SMS-WSJ is in the range of \SIrange{5}{6}{\decibel} \gls{siSDR} for single-channel input and single-stage processing \cite{21_Wang_complex_mapping, 20_Maciejewski_WHAMR,21_Zheghidour_wavesplit}, which is much worse than the performance on clean, anechoic mixtures. In this contribution, we employ SMS-WSJ for our experiments because we wish to assess the performance of the separation system not only by the signal-related evaluation metric \gls{SDR} but also by \gls{WER}, given that the SMS-WSJ Kaldi recipe allows us to compare the \gls{WER} performance across different publications.

This paper is not about suggesting a new algorithm for reverberant source separation. We rather aim to explore, in a systematic way, which of the recent innovations that proved useful for the separation of anechoic mixtures are also beneficial in the reverberant case, in order to propose some guidelines on how to adjust a separation system to reverberated input.

As our outset, we take the SepFormer architecture, which achieves state-of-the-art performance both on \gls{WSJ0-2mix} \cite{Subakan_21_sepformer} and WHAMR! \cite{22_Sepformer_refs}, and the traditional \gls{PIT-BLSTM} source separation model from \cite{Kolbaek_17_uPit}.
Here, we modify and optimize the \gls{PIT-BLSTM} to detect which differences between both models aside from the separator lead to a better separation performance.
Then, we modify the SepFormer \wrt loss function, encoder/decoder architecture and resolution to mitigate the performance degradation between the anechoic and reverberant scenario. Indeed, we are able to improve the performance \wrt \gls{WER} by \num{7} percentage points compared to the vanilla SepFormer implementation. Nevertheless, the final result turns out to be hardly superior to that of the optimized \gls{PIT-BLSTM}, calling into question  the importance of some of the innovations of recent years for the realistic case of reverberant speech separation.

The remainder of the paper is structured as follows. In \cref{sec:separation_pipeline} the \gls{PIT-BLSTM} and the SepFormer are briefly introduced as two realizations 
of an
abstracted pipeline for mask-based source separation. \Cref{sec:assumptions} discusses design choices in light of the requirements of a reverberated input.  In
\Cref{sec:evaluation} the SepFormer is optimized for performance on reverberant data and compared to the PIT model in \cref{sec:comparison}.  The paper concludes
with a short discussion in \cref{sec:conc}.

\section{Mask-based source separation}
\label{sec:separation_pipeline}
\begin{figure}
    \centering
    \input{images/masking_pipeline.tex}
    \caption{Block diagram of mask-based source separation
    }
    \label{fig:separation_pipeline}
\end{figure}
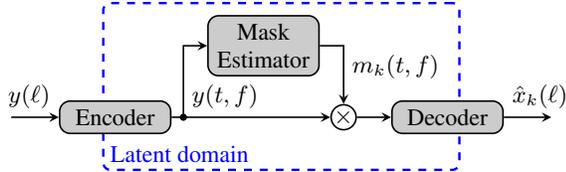

Mask-based systems for single-channel source separation can be 
abstracted to the same general processing pipeline that is  depicted in \cref{fig:separation_pipeline}.
First, the observed time-domain signal $y(\ell)$ is transformed into a latent space (e.g., the \gls{STFT} domain or a learned representation). 
In this latent space, the encoded mixture $y(t,f)$ with time index $t$ and latent feature index $f$ is used as the input of the neural separation module, which estimates a mask $m_k(t,f)$ for the reconstruction of each active speaker $k$ in the observation. Then, the estimated signal $\hat{x}_k(t,f)$ of each speaker is obtained by masking the mixture with the estimated masks
\begin{align}
    \label{eq:mask_multiplication}
    \hat{x}_k(t,f) = y(t,f)m_k(t,f).
\end{align}
The reconstructed signals $\hat{x}_k(t,f)$ are then transformed back into the time-domain in the decoder.

Both the \gls{PIT-BLSTM} approach to monaural source separation \cite{Kolbaek_17_uPit} and the SepFormer \cite{Subakan_21_sepformer} (the latter providing state-of-the-art results on \gls{WSJ0-2mix}) use a mask-based separation.
By comparing these two models, which, in principle, share the same overall structure of \cref{fig:separation_pipeline}, we investigate if modifications that were found to be useful in the anechoic scenario can be transferred to the reverberant case.

\section{Source separation under reverberation}
\label{sec:assumptions}
Mask-based source separation relies on the sparsity and orthogonality of the sources in the the domain where the masks are computed. 
In case of the \gls{STFT} domain, this means that a time-frequency bin $(t,f)$ of a mixture $y(t,f)$ can be approximated by the contribution of the dominant source $i(t,f)$ 

\begin{align}
    y(t,f) &= \sum_{k=1}^K s_k(t,f)h_k(t,f) \nonumber \\
            &\approx s_{i(t,f)}(t,f)h_{i(t,f)}(t,f)
            \label{eq_orth_mtfa}
\end{align}
where $s_k(t,f)$ and $h_k(t,f)$ are the STFT representations of the $k$-th source signal and the \gls{RIR} from the $k$-th source to the microphone, respectively. Further, $i(t,f)\in\{1,\ldots K\}$ 
indicates which of the $K$ sources dominates in bin $(t,f)$.





Note that Eq.~\eqref{eq_orth_mtfa} makes the additional  assumption that the convolution of the source signal $s(\ell)$ with the \gls{RIR} $h(\ell)$  corresponds to a multiplication of their respective STFT transforms.
This so-called \gls{MTFA}, however, only holds true if the temporal extent of $h(\ell)$ is smaller than the STFT analysis window  \cite{07_Avargel_mtfa}. 
When  the window length is decreased, this assumption becomes more and more questionable,  and the \gls{CTFA} \cite{09_Talmon_ctfa} would be more appropriate.
Obviously, this challenges mask-based source reconstruction according to Eq.~\eqref{eq:mask_multiplication}, and the complications are the more pronounced the smaller the STFT analysis window is. 

When switching from a fixed STFT encoder to a learnable encoder, the overall structure of the system, see \Cref{fig:separation_pipeline}, stays the same. Therefore, it can be assumed that similar issues arise with the learnable encoder.
In the following we will thus study the influence of the encoder/decoder and their temporal resolution on the separation performance.

\section{Evaluation}
\label{sec:evaluation}

\subsection{Database and Baseline Results}
In order to assess which effect a specific component of a separation module has both on nonreverberant and reverberant data, it is important to run the experiments on a corpus that differs only in this respect. We employ the  SMS-WSJ data set \cite{Drude_19_smswsj} for our analysis, which easily allows us to generate both anechoic and reverberant two-speaker mixtures.

For the anechoic scenario, the 
reverberation time $T_{60}$ is reduced from \SIrange{0.2}{0.5}{\second} to zero
while keeping an otherwise identical data simulation.
Dynamic mixing is employed in training: each example during training consists of randomly drawn utterances from WSJ database and only the RIRs are reused to provide a dramatically increased number of examples, which has been proven to improve the system's performance \cite{21_Zheghidour_wavesplit}.
To show the competitiveness of the used models, we also provide baseline results on the \gls{WSJ0-2mix} \cite{16_Hershey_dc_wsjmix} database.

\begin{table}[tb]
\vspace{-6pt}  
\caption{Performance of the baseline models on the (anechoic) \gls{WSJ0-2mix} database}
 
\label{tab:sdr_ref_wsj2mix}
    \centering
    \begin{tabular}{l S S}
    \toprule
         Model &  {SDR} & {\#Params} \\
         \midrule
         SepFormer \cite{Subakan_21_sepformer} & 20.4 & 25.7M \\
         SepFormer (small) & 19.3 & 13.0M\\
         \gls{PIT-BLSTM} \cite{Kolbaek_17_uPit} & 9.8 & 23.5 M\\
         \bottomrule
    \end{tabular}
\end{table}

\begin{table}
\vspace{-6pt}  
\caption{
\gls{SDR} of the baseline models on anechoic and reverberant SMS-WSJ data on the test dataset with matched training data
}
\label{tab:sdr_ref_smswsj}
    \centering
    \begin{tabular}{l S S S S}
    \toprule
         Model &  \multicolumn{2}{c}{anechoic } & \multicolumn{2}{c}{reverb} \\
                  \cmidrule(lr){2-3}  \cmidrule(lr){4-5}
         & {SDR} & {WER} & {SDR} & {WER} \\ 
         \midrule
         \gls{PIT-BLSTM} & 10.27 & 39.81  &   7.77  & 52.78  \\
         SepFormer (small) & \bfseries 19.13 & 13.14 & 8.98 & 41.43 \\
         \bottomrule
    \end{tabular}
\end{table}

The \gls{PIT-BLSTM} model consists of 3 BLSTM layers with 600 units each, followed by 2 fully connected layers.
The encoder and decoder are set to the \gls{STFT} and inverse \gls{STFT} with a window size of \num{512}, a frame advance of \num{128} and an embedding dimension (number of frequency bins) of \num{257} at \SI{8}{\kilo\hertz} sampling rate. The output of the STFT encoder is the concatenated real and imaginary part of the spectrum as in \cite{20_Heitkaemper_dissecting}.

The SepFormer uses the same parameters as 
proposed in 
\cite{Subakan_21_sepformer} with a window size of \num{16}, a frame advance of \num{8} and a latent dimension of \num{256}, aside from reducing the number of intra- and  inter-Transformer layers to \num{4}, each. 
This modification yields an about \SI{1}{\decibel} lower \gls{SDR} on WSJ0-2mix, but
significantly reduces the number of parameters, see entry \enquote{SepFormer (small)} in \cref{tab:sdr_ref_wsj2mix}. Thus, for all following experiments this \enquote{small} configuration is employed due to computational limitations. Note, that the memory footprint of the small SepFormer still is \num{16} times larger than the \gls{PIT-BLSTM}, so that a complexity comparison purely based on the parameters is not fair.
The learnable encoder is a single CNN layer with \num{256} channels, i.e. the latent size, followed by a ReLU, and the decoder has only one CNN layer as in \cite{19_Luo_convtasnet}.

Both architectures use 
the Adam optimizer \cite{05_Kingma_adam} and the early reverberated signals as target as proposed in \cite{Drude_19_smswsj}.
The SepFormer is trained with a soft-thresholded time-domain
\gls{SDR} loss \cite{20_Wisdom_mixtures_of_mixtures} 
\begin{align}
    \mathcal{L}^{\mathrm{th-SDR}} = 
    10\log_{10} \frac{1}{K} \sum_k \left( \frac{\sum\limits_\ell \left\lvert\hat{x}_{k}(\ell) - x_k(\ell)\right\rvert^2}{\sum\limits_\ell \left\lvert x_k(\ell)\right\rvert^2} + \tau\right),
    \label{eq:th_sdr}
\end{align}
where $\tau=10^{-\mathrm{SDRmax}/10}$ and $\mathrm{SDRmax} =  \text{\SI{20}{\decibel}} $.
This loss decreases the contribution of well separated examples to the gradient, encouraging the model to focus more on enhancing examples with a low \gls{SDR} than those that already show a good separation. 
The Baseline PIT-BLSTM is trained with a frequency-domain \gls{SDR} loss.
The models are evaluated \wrt \gls{SDR}, PESQ \cite{01_rix_pesq}, and \gls{WER}.
We use the \gls{SDR} metric proposed in \cite{06_Vincent_sdr}, as it allows an evaluation against the anechoic speech source. The PESQ values also are given \wrt the speech source, and 
the \gls{WER} results on SMS-WSJ are determined with the acoustic model from \cite{Drude_19_smswsj}.

\Cref{tab:sdr_ref_wsj2mix} and  \cref{tab:sdr_ref_smswsj} display the results of the baseline systems \cite{Kolbaek_17_uPit, Subakan_21_sepformer} on WSJ0-2mix and SMS-WSJ, respectively.
It can be seen that both systems degrade under the presence of reverberation.
However, the separation performance of the SepFormer degrades by more than \SI{10}{\decibel} in terms of \gls{SDR} and almost \num{30} percentage points regarding the \gls{WER}. 
We wish to find out which components of the SepFormer make it become so sensitive to reverberation.
\subsection{PIT-BLSTM optimization}
First, we optimized the performance of  the \gls{PIT-BLSTM} on reverberant input data.
To do so, we  switched the training objective from the frequency-domain loss to the thesholded time-domain loss described in \cref{eq:th_sdr}. In this way, even though the \gls{PIT-BLSTM} uses the magnitude spectrum for the mask estimation, the phase has an influence on the computed loss.
In addition, we added white Gaussian noise at an SNR of \SI{25}{\decibel} to the separated audio files before they were input to the speech recognizer. This is to mask artefacts that were introduced during the source separation. 
Next to an improved \gls{WER} we also observed a higher correlation between the signal-level metric \gls{SDR} and the \gls{WER}, rendering the \gls{SDR} a better predictor of the ASR performance.
As shown in \cref{tab:reference_sep}, by introducing the latter modifications
 the performance of the \gls{PIT-BLSTM} is significantly improved both in terms of \gls{SDR} and \gls{WER}. Even more so, these modifications work well both with and without reverberation and lead to a reduction in \gls{WER} of more than 20 percentage points for both scenarios.

\begin{table}[bt]
	\caption{Comparison of the optimzed PIT-BLSTM and the baseline SepFormer model on SMS-WSJ }
	\label{tab:reference_sep}
	\sisetup{detect-weight}
	\robustify\bfseries  
	\sisetup{detect-weight=true,detect-inline-weight=text}
	\centering
	\begin{tabular}{l S S S S}
		\toprule
		Model &  \multicolumn{2}{c}{anechoic } & \multicolumn{2}{c}{reverb} \\
		\cmidrule(lr){2-3}  \cmidrule(lr){4-5}
		& {SDR} & {WER} & {SDR} & {WER} \\ 
		\midrule
		PIT-BLSTM & 10.27 & 39.81  &   7.77  & 52.78  \\
		PIT-BLSTM (th-SDR)  & 14.13 & 19.65 & \bfseries 10.93  &  35.70 \\  
		\quad + Gaussian noise & {-} & 13.19 & {-}  & \bfseries 27.47 \\    
		SepFormer (small) & \bfseries 19.13 & 13.14 & 8.98 & 41.43 \\
		\quad + Gaussian noise & {-} & \bfseries 9.57 & {-} & 33.51 \\
		\bottomrule
	\end{tabular}
\end{table}

\subsection{SepFormer optimization}
The above changes to the \gls{PIT-BLSTM} system also lead to improvements of the Sepformer, see \cref{tab:reference_sep}. Therefore, the Gaussian noise is added in all following evaluations.
However, it is striking that the Sepformer is no longer superior to the \gls{PIT-BLSTM} system for reverberant data.
Therefore, we gradually exchanged the Sepformer's components with those of the \gls{PIT-BLSTM} system to investigate the cause of this performance loss and what the best configuration is for reverberant input.


\subsubsection{Encoder/decoder choice}
There is a large mismatch between the window size and the frame advance of standard PIT-BLSTM and SepFormer systems.
 To verify whether the violation of the \gls{MTFA} caused by the small window size of the SepFormer contributes to the system deterioration under reverberation, we evaluated the SepFormer for multiple encoder/decoder configurations. As opposed to other works \cite{20_Heitkaemper_dissecting}, we only increase the window size while maintaining small shift sizes in order to retain a high temporal resolution. 
 \Cref{tab:encoder_sizes_anechoic_and_reverberant} shows the expected behavior for the SepFormer in anechoic conditions: reducing the frame shift leads to an improvement in \gls{SDR} and \gls{WER}. The recommended analysis window size and shift of 16 and 8 samples (i.e. \SI{2}{\milli\second} and \SI{1}{\milli\second}) \cite{Subakan_21_sepformer}, respectively, provides the best results for anechoic data.
 Furthermore, the learnable encoder proves superior to the \gls{STFT} encoder.

\input{table_learnable_vs_stft.tex}
 
 Conversely, for the reverberant scenario,
 while the STFT encoder in \cref{tab:encoder_sizes_anechoic_and_reverberant} is significantly worse than a learnable encoder for a small window size and shift, it begins to be on par or even outperforms the learnable encoder for an increased window size  of \SI{32}{\milli\second}.
 This validates our assumption that the violation of the \gls{MTFA} contributes to the poor model performance under reverberation.
 Interestingly, the overall best results of the SepFormer are achieved 
 with the STFT. 


Our assumptions are further supported by the \gls{WDO} \cite{02_Rickard_WDO} score which measures the orthogonality of the single-speaker utterances in the latent space. 
Following on the results from \cref{tab:encoder_sizes_anechoic_and_reverberant}
it becomes apparent that the baseline SepFormer learns a highly orthogonal space for anechoic data. However, by switching to reverberant data, the WDO decreases by 5 percentage points.  This decrease is mitigated by a larger encoder window size. The same is true for the STFT encoder, where the regularizing effect of a larger window size is even more pronounced. This indicates that the learnable encoder is able to compensate the effects to some degree, but that choosing a large enough window size is mandatory to stabilize the performance under reverberation.

\subsubsection{Data representation for the mask estimator}
\label{sec:phase_realimag}
A significant difference between \gls{PIT-BLSTM} and SepFormer is that the PIT model estimates the masks based on the magnitude spectrogram  only, whereas the SepFormer mask estimator has access to the complete signal, i.e., both magnitude and phase in case of the STFT representation.

\begin{table}[bt]
    \centering
    \sisetup{detect-weight}
	\robustify\bfseries  
	\sisetup{detect-weight=true,detect-inline-weight=text}
    \vspace{-6pt}  
    \caption{Separation performance of the SepFormer for different input representations of the STFT features on reverberant SMS-WSJ}
    \label{tab:phase_info}
    \addtolength{\tabcolsep}{-0.05em}   
    \begin{tabular}{c c c c S S S}
        \toprule
        \multirow{2}{*}{\shortstack[c]{win. \\ size}} & \multirow{2}{*}{\shortstack[c]{latent \\ size}} & shift & Input data & {SDR} & {PESQ}& {WER}   \\
        \\
        \midrule
        256 & 256 & 16 & Real+Imag & 10.79 & 1.90 & 29.10 \\ 
        256 & 256 & 64 & Real+Imag & 10.01 & 1.83 & 31.80 \\ 
        512 & 256 & 16 & Magnitude & 10.48 & 1.82 & 29.22 \\        
        512 & 256 & 128 & Magnitude & \bfseries 11.00 &  \bfseries 1.91  & \bfseries 26.50  \\    
        \bottomrule
    \end{tabular}
\end{table}

To compare both networks with the same input representation, the effect of only using the magnitude as input for the mask estimator in the SepFormer is evaluated. 
The SepFormer trained with concatenated real and imaginary parts estimates separate masks for the real and imaginary parts of the observation, respectively. When only using the magnitude for the mask estimation, the estimated masks are applied both on the real and imaginary parts.
\Cref{tab:phase_info} 
shows that the availability of the phase information is not helpful for the SepFormer in the reverberant scenario. Even more so, omitting the phase information leads to a better system performance.

This can have two reasons. Firstly, only using the magnitude spectrogram results in a larger window of \num{512} samples to keep the size of the separator identical, increasing the temporal context of each frame even further. Secondly, \cite{Peer_22_phase_enhancement}  has shown that the phase becomes less informative while the magnitude becomes more informative for increasing frame sizes. The configurations trained with both the phase and magnitude information learn a trade-off between phase and magnitude reconstruction. However, the large window sizes that were shown to be necessary in \cref{tab:encoder_sizes_anechoic_and_reverberant} for the reverberant scenario result in an uninformative phase representation.
Therefore, omitting this information only slightly deteriorates the system performance for a small frame shift.
However, by further increasing the frame shift the magnitude spectrogram becomes more informative. 
Therefore, using the magnitude allows increasing the frame shift from 16 to 128 samples, reducing the computational effort by almost a factor of \num{8} compared to the best configuration in \cref{tab:encoder_sizes_anechoic_and_reverberant} while simultaneously improving both  signal-level metrics and \gls{WER}. 

\section{Summary}
\label{sec:comparison}
\begin{table}[bt]
    \centering
    \sisetup{detect-weight}
	\robustify\bfseries  
	\sisetup{detect-weight=true,detect-inline-weight=text}
    \vspace{-6pt}  
	\caption{Performance comparison of  the best anechoic and reverberant system configurations}
    \label{tab:summary_sepformer_pit}
    \begin{tabular}{l S S S S}
    \toprule
    System & \multicolumn{2}{c}{anechoic} & \multicolumn{2}{c}{reverb}\\
    \cmidrule(lr){2-3}\cmidrule(lr){4-5}
     & {SDR} & {WER} & {SDR} & {WER} \\
    \toprule
    opt. \gls{PIT-BLSTM} & 14.13 & 13.19 & 10.93 & 27.47\\
    opt. SepFormer anechoic & 19.13 &9.57 & 8.98 & 33.51\\
    opt. SepFormer reverb & 14.03 & 14.09 & 11.00 &26.50\\
    \bottomrule
\end{tabular}
\end{table}

\Cref{tab:summary_sepformer_pit} summarizes the performance of the SepFormer on anechoic and reverberant SMS-WSJ using the best configuration for anechoic data as reported in \cite{Subakan_21_sepformer} and the best configuration for reverberated input as found here, and compares it with the performance of the optimized PIT-BLSTM system. Interestingly, the SepFormer configuration that was found optimal for reverberant input is quite similar to the PIT-BLSTM: it uses a fixed STFT encoder with the magnitude spectrogram at its input and the same window size and frame shift. Only the network architecture of the separator is different, i.e., intra- and inter-transformer layers vs BLSTM layers. However, this modified SepFormer only shows a marginally better \gls{SDR} and an improvement of \num{1} percentage point in the \gls{WER}.


\section{Conclusions}
\label{sec:conc}



In this paper, we investigated the impact of reverberation on the various design choices for the SepFormer  source separation system that is considered state-of-the-art for anechoic mixtures.
We showed that it is mandatory  to choose a large enough encoder window size for reverberant data. Also, we demonstrated that the STFT likewise is a valid choice as encoder and decoder. Here, it becomes apparent that the phase information no longer is helpful for the separation and only using the magnitude information provides superior results while reducing the computational complexity significantly.

Despite several modifications which greatly improved the performance of the SepFormer on reverberated mixtures, it was in the end hardly superior to a  \gls{PIT-BLSTM} separation system,  
which was optimized with only rather straightforward modifications, such as loss computation in time domain.  At least for a single-stage approach, our experiments indicate that jointly focusing on phase and magnitude reconstruction leads to subpar results compared to solely focusing on magnitude reconstruction under reverberation.
This raises the issue of whether the improvements that have been appraised for the separation of anechoic mixtures, such as learnable encoder and phase reconstruction, are futile for the more realistic case of reverberant source separation.

We therefore argue that research on source separation should primarily focus on the practically more relevant case of reverberant input, rather than on the anechoic scenario. Since jointly tackling both reverberation and overlapped speech appears to be a challenging task, an alternative solution is to apply an explicit dereverberation component and/or employ multi-stage processing, as in \cite{Wang_21_sms_twostage}. 


\section{Acknowledgement}
Computational resources were provided by the Paderborn Center for Parallel Computing.
C. Boeddeker was supported by DFG under project no. 448568305.

\vfill\pagebreak

\bibliographystyle{IEEEbib}
\bibliography{refs}

\end{document}

%% file: images/masking_pipeline.tex

\begin{tikzpicture}[semithick,auto,
block/.style={
		rectangle,
		draw,
		fill=black!20,
		text centered,
		text width=4em,
		rounded corners,
		minimum width=4em},
mul/.style={
        circle,
        draw,
    },		
	]
\tikzset{>=stealth}
\tikzstyle{branch}=[{circle,inner sep=0pt,minimum size=0.3em,fill=black}]

\tikzset{pics/.cd,
	pic switch closer/.style args={#1 times #2}{code={
		\tikzset{x=#1/2,y=#2/2}
		\coordinate (-in) at (1,0);
		\coordinate (-out) at (-1,0);
		
		\draw [line cap=rect] (-1, 0) -- ++(0.1,0) -- ++(20:1.9);
		
	}}
}

    \pgfdeclarelayer{background}
    \pgfdeclarelayer{foreground}
    \pgfsetlayers{background,main,foreground}

    \node [block] (decoder) {Decoder};
    \node [block, anchor=south east] (separator) at ($(decoder.north west) + (-3em, 1em)$) {Mask Estimator};
    \node [circle, draw=black, inner sep=0cm] (masking) at ($(decoder.west) + (-2em, 0)$) {$\times$};
    \node [block, anchor=east] (encoder) at ($(separator.west |- decoder) + (-1.5em, 0)$) {Encoder};
    
    \draw[<-] (encoder.west) node[above left] {$y(\ell)$} -- +(-2em, 0);
    \draw[->] (decoder.east) node[above right] {$\hat{x}_k(\ell)$} -- +(2em, 0);
    \draw[->] (encoder.east) -- (masking);
    \draw[->] ($(encoder.east)!1/3!(encoder.east-|separator.west)$) node[branch]{} node[above right] {$y(t,f)$} |- (separator.west);
    \draw[->] (separator.east) -- (separator-|masking) node[below right] {$m_k(t,f)$} -- (masking);
    \draw[->] (masking) -- (decoder);

    


    

    
\begin{pgfonlayer}{background}
	\coordinate(tmp) at ($(encoder.south east) + (0, -1em)$);
    \tikzstyle{box} = [draw, dashed, inner xsep=0.5em, inner ysep=0.5em, line width=0.1em, rounded corners=0.3em]
	\node (box) [box, blue, text=blue, fit=(encoder.center) (separator.north) (decoder.center) (tmp), label={[anchor=south west, align=left]south west:\color{blue}Latent domain}] {};
    
\end{pgfonlayer}



\end{tikzpicture}

%% file: table_learnable_vs_stft.tex
\begin{table*}[hbt]
    \sisetup{detect-weight}
	\robustify\bfseries  
    \sisetup{round-precision=2,round-mode=places, table-format = 2.1}
%
    \centering
    \vspace{-6pt}  
        \caption{Separation performance of the SepFormer on SMS-WSJ with a learnable and STFT encoder/decoder and varying  encoder shifts/sizes}
    \label{tab:encoder_sizes_anechoic_and_reverberant}
    \begin{tabular}{c c c c S S S S S S S S}
    \toprule
    \multirow{2}{*}{\shortstack[c]{win. \\ size}} & \multirow{2}{*}{\shortstack[c]{latent \\ size}} & \multirow{2}{*}{shift} & \multirow{2}{*}{\shortstack[c]{learnable \\ encoder}} & 
        \multicolumn{4}{c}{anechoic data} & \multicolumn{4}{c}{reverberant data} \\
    \cmidrule(lr){5-8}\cmidrule(lr){9-11}
     &  &  &  & 
        {SDR [\si{\decibel}]} & {WDO [\si{\percent}]}&{PESQ} & {WER [\si{\percent}]} & {SDR [\si{\decibel}]} &  {WDO [\si{\percent}]}& {PESQ} &{WER [\si{\percent}]} \\
    \midrule
    16  & 256  & 8 & \cmark & 
        \bfseries 19.13  & \bfseries 85.483895 &\bfseries 3.43  & \bfseries 9.57  &   
        8.98 & 79.9359321594 & 1.83 & 33.51 \\           
    256 & 256  & 8 & \cmark & 
        16.68 & 82.759815 &3.11 & 12.28 &  
        10.56 & 82.88679718971252&\bfseries 1.91 & 31.75 \\     
    256 & 256  & 16 & \cmark & 
        15.27 & 81.88877105712891 & 3.00 &  13.96 &   
        10.23 & 81.94860219955444 &1.85 & 30.66 \\        
    256 & 256  & 64 & \cmark & 
        11.86 & 84.26356315612793 &  2.47 & 20.54 &  
        9.54 & \bfseries 85.07705926895142 & 1.83 & 34.83\\         
 \midrule
    16  & 256  & 8 & \xmark & 
        16.74 & 73.58398986230643 & 2.84 & 11.90 &  
        7.44 & 69.95357139360081& 1.71 & 45.83\\           
    256 & 256  & 8 & \xmark & 
        15.70 & 79.24350439190095 & 2.91 & 11.61&   
        9.97 & 77.5708516581603  & 1.84 & 31.38\\           
    256 & 256  & 16 & \xmark & 
        14.47 & 79.24365930319872 & 2.69 & 13.52 & 
        \bfseries 10.79 & 77.5707768761923 & 1.90 & \bfseries 29.10  \\   
    256 & 256  & 64 & \xmark & 
        13.52 & 79.24355352549952 & 2.66 &15.02 &  
        10.01 & 77.5725939190082 & 1.83 & 31.80 \\        
    \bottomrule
    \end{tabular}

\end{table*}


%% file: main.bbl
\begin{thebibliography}{10}

\bibitem{16_Hershey_dc_wsjmix}
John~R. Hershey, Zhuo Chen, Jonathan Le~Roux, and Shinji Watanabe,
\newblock ``Deep clustering: Discriminative embeddings for segmentation and
  separation,''
\newblock in {\em IEEE International Conference on Acoustics, Speech and Signal
  Processing (ICASSP)}, 2016, pp. 31--35.

\bibitem{Kolbaek_17_uPit}
Morten Kolbæk, Dong Yu, Zheng-Hua Tan, and Jesper Jensen,
\newblock ``Multitalker speech separation with utterance-level permutation
  invariant training of deep recurrent neural networks,''
\newblock {\em IEEE/ACM Transactions on Audio, Speech, and Language
  Processing}, vol. 25, no. 10, pp. 1901--1913, 2017.

\bibitem{17_Luo_chimera}
Yi~Luo, Zhuo Chen, John~R. Hershey, Jonathan Le~Roux, and Nima Mesgarani,
\newblock ``Deep clustering and conventional networks for music separation:
  Stronger together,''
\newblock in {\em IEEE International Conference on Acoustics, Speech and Signal
  Processing (ICASSP)}, 2017, pp. 61--65.

\bibitem{18_Luo_tasnet}
Yi~Luo and Nima Mesgarani,
\newblock ``{TasNet}: Time-domain audio separation network for real-time,
  single-channel speech separation,''
\newblock in {\em IEEE International Conference on Acoustics, Speech and Signal
  Processing (ICASSP)}, 2018, pp. 696--700.

\bibitem{19_Luo_convtasnet}
Yi~Luo and Nima Mesgarani,
\newblock ``{Conv-TasNet}: Surpassing ideal time–frequency magnitude masking
  for speech separation,''
\newblock {\em IEEE/ACM Transactions on Audio, Speech, and Language
  Processing}, vol. 27, no. 8, pp. 1256--1266, 2019.

\bibitem{20_Luo_dprnn}
Yi~Luo, Zhuo Chen, and Takuya Yoshioka,
\newblock ``{Dual-Path RNN}: Efficient long sequence modeling for time-domain
  single-channel speech separation,''
\newblock in {\em IEEE International Conference on Acoustics, Speech and Signal
  Processing (ICASSP)}, 2020, pp. 46--50.

\bibitem{Subakan_21_sepformer}
Cem Subakan, Mirco Ravanelli, Samuele Cornell, Mirko Bronzi, and Jianyuan
  Zhong,
\newblock ``Attention is all you need in speech separation,''
\newblock in {\em IEEE International Conference on Acoustics, Speech and Signal
  Processing (ICASSP)}, 2021, pp. 21--25.

\bibitem{20_Maciejewski_WHAMR}
Matthew Maciejewski, Gordon Wichern, Emmett McQuinn, and Jonathan Le~Roux,
\newblock ``{WHAMR!}: Noisy and reverberant single-channel speech separation,''
\newblock in {\em IEEE International Conference on Acoustics, Speech and Signal
  Processing (ICASSP)}, 2020, pp. 696--700.

\bibitem{Drude_19_smswsj}
Lukas Drude, Jens Heitkaemper, Christoph Boeddeker, and Reinhold Haeb-Umbach,
\newblock ``{SMS-WSJ}: Database, performance measures, and baseline recipe for
  multi-channel source separation and recognition,''
\newblock {\em arXiv preprint arXiv:1910.13934}, 2019.

\bibitem{11_Povey_kaldi}
Daniel Povey, Arnab Ghoshal, Gilles Boulianne, Nagendra Goel, Mirko Hannemann,
  Yanmin Qian, Petr Schwarz, and Georg Stemmer,
\newblock ``The {Kaldi} speech recognition toolkit,''
\newblock in {\em IEEE Workshop on Automatic Speech Recognition and
  Understanding (ASRU)}, 2011.

\bibitem{21_Wang_complex_mapping}
Zhong-Qiu Wang, Peidong Wang, and DeLiang Wang,
\newblock ``Multi-microphone complex spectral mapping for utterance-wise and
  continuous speech separation,''
\newblock {\em IEEE/ACM Transactions on Audio, Speech, and Language
  Processing}, vol. 29, pp. 2001--2014, 2021.

\bibitem{21_Zheghidour_wavesplit}
Neil Zeghidour and David Grangier,
\newblock ``Wavesplit: {End-to-End} speech separation by speaker clustering,''
\newblock {\em IEEE/ACM Transactions on Audio, Speech, and Language
  Processing}, vol. 29, pp. 2840--2849, 2021.

\bibitem{22_Sepformer_refs}
Cem Subakan, Mirco Ravanelli, Samuele Cornell, Francois Grondin, and Mirko
  Bronzi,
\newblock ``On using transformers for speech-separation,''
\newblock {\em arXiv preprint arXiv:2202.02884}, 2022.

\bibitem{07_Avargel_mtfa}
Yekutiel Avargel and Israel Cohen,
\newblock ``On multiplicative transfer function approximation in the short-time
  fourier transform domain,''
\newblock {\em IEEE Signal Processing Letters}, vol. 14, no. 5, pp. 337--340,
  2007.

\bibitem{09_Talmon_ctfa}
Ronen Talmon, Israel Cohen, and Sharon Gannot,
\newblock ``Relative transfer function identification using convolutive
  transfer function approximation,''
\newblock {\em IEEE/ACM Transactions on Audio, Speech, and Language
  Processing}, vol. 17, no. 4, pp. 546--555, 2009.

\bibitem{20_Heitkaemper_dissecting}
Jens Heitkaemper, Darius Jakobeit, Christoph Boeddeker, Lukas Drude, and
  Reinhold Haeb-Umbach,
\newblock ``Demystifying {TasNet}: A dissecting approach,''
\newblock in {\em IEEE International Conference on Acoustics, Speech and Signal
  Processing (ICASSP)}, 2020, pp. 6359--6363.

\bibitem{05_Kingma_adam}
Diederik~P. Kingma and Jimmy Ba,
\newblock ``Adam: {A} method for stochastic optimization,''
\newblock in {\em International Conference on Learning Representations (ICLR)},
  2015.

\bibitem{20_Wisdom_mixtures_of_mixtures}
Scott Wisdom, Efthymios Tzinis, Hakan Erdogan, Ron Weiss, Kevin Wilson, and
  John Hershey,
\newblock ``Unsupervised sound separation using mixture invariant training,''
\newblock in {\em Advances in Neural Information Processing Systems}, 2020,
  vol.~33, pp. 3846--3857.

\bibitem{01_rix_pesq}
A.W. Rix, J.G. Beerends, M.P. Hollier, and A.P. Hekstra,
\newblock ``Perceptual evaluation of speech quality (pesq)-a new method for
  speech quality assessment of telephone networks and codecs,''
\newblock in {\em 2001 IEEE International Conference on Acoustics, Speech, and
  Signal Processing. Proceedings (Cat. No.01CH37221)}, 2001, vol.~2, pp.
  749--752 vol.2.

\bibitem{06_Vincent_sdr}
Emmanuel Vincent, R{\'e}mi Gribonval, and C{\'e}dric F{\'e}votte,
\newblock ``{Performance measurement in blind audio source separation},''
\newblock {\em IEEE/ACM Transactions on Audio, Speech and Language Processing},
  vol. 14, no. 4, pp. 1462--1469, 2006.

\bibitem{02_Rickard_WDO}
Scott Rickard and Ozgiir Yilmaz,
\newblock ``On the approximate w-disjoint orthogonality of speech,''
\newblock in {\em 2002 IEEE International Conference on Acoustics, Speech, and
  Signal Processing}, 2002, vol.~1, pp. I--529--I--532.

\bibitem{Peer_22_phase_enhancement}
Tal Peer and Timo Gerkmann,
\newblock ``Phase-aware deep speech enhancement: It's all about the frame
  length,''
\newblock {\em arXiv preprint arXiv:2203.16222}, 2022.

\bibitem{Wang_21_sms_twostage}
Zhong-Qiu Wang, Gordon Wichern, and Jonathan~Le Roux,
\newblock ``Convolutive prediction for monaural speech dereverberation and
  noisy-reverberant speaker separation,''
\newblock {\em IEEE/ACM Transactions on Audio, Speech, and Language
  Processing}, vol. 29, pp. 3476--3490, 2021.

\end{thebibliography}
